\documentclass{aa}  

\usepackage{graphicx}
\usepackage{txfonts}
\usepackage{xcolor}
\usepackage{comment}
\usepackage{natbib,twoopt}
\usepackage[breaklinks=true]{hyperref} %% to avoid \citeads line fills
\bibpunct{(}{)}{;}{a}{}{,} %% natbib format for A&A and ApJ
\makeatletter
\newcommandtwoopt{\citeads}[3][][]{\href{http://adsabs.harvard.edu/abs/#3}%
{\def\hyper@linkstart##1##2{}%
\let\hyper@linkend\@empty\citealp[#1][#2]{#3}}}
\newcommandtwoopt{\citepads}[3][][]{\href{http://adsabs.harvard.edu/abs/#3}%
{\def\hyper@linkstart##1##2{}%
\let\hyper@linkend\@empty\citep[#1][#2]{#3}}}
\newcommandtwoopt{\citetads}[3][][]{\href{http://adsabs.harvard.edu/abs/#3}%
{\def\hyper@linkstart##1##2{}%
\let\hyper@linkend\@empty\citet[#1][#2]{#3}}}
\newcommandtwoopt{\citeyearads}[3][][]%
{\href{http://adsabs.harvard.edu/abs/#3}
{\def\hyper@linkstart##1##2{}%
\let\hyper@linkend\@empty\citeyear[#1][#2]{#3}}}
\makeatother

\newcommand{\kms}{\mbox{\,km\,s$^{-1}$}}
\newcommand{\msun}{\mbox{\,M$_\odot$}}

\begin{document}

   \title{The rich molecular environment of the luminous blue variable star \mbox{AFGL\,2298}
   \thanks{Based on observations carried out with the IRAM-30\,m telescope. The Institut de Radioastronomie Millim\'etrique (IRAM) is supported by INSU/CNRS (France), MPG (Germany), and IGN (Spain).}$^,$\thanks{Reduced spectra, data cubes and model results are available at the CDS via \url{https://cdsarc.cds.unistra.fr/viz-bin/cat/J/A+A/xxx/Ayyy}}
   }

   \author{J. R. Rizzo
          \inst{1,2}
          \and
          C. Bordiu\inst{3}
          \and
          C. Buemi\inst{3}
          \and
          P. Leto\inst{3}
          \and
          A. Ingallinera\inst{3}
          \and
          F. Bufano\inst{3}
          \and
          G. Umana\inst{3}
          \and
          L. Cerrigone\inst{4}
          \and
          C. Trigilio\inst{3}
          }

   \institute{
              {ISDEFE, Beatriz de Bobadilla 3, E-28040 Madrid, Spain}
         \and {Centro de Astrobiolog\'{\i}a (CSIC-INTA), Ctra.~M-108, km.~4, 
               E-28850 Torrej\'on de Ardoz, Spain}\\
              \email{ricardo.rizzo@cab.inta-csic.es}
         \and {INAF--Osservatorio Astrofisico di Catania, Via Santa Sofia 78, I-95123 Catania, Italy}
         \and {Joint ALMA Observatory, Alonso de Córdova 3107, Vitacura, 8320000, Santiago, Chile}
}

   \date{Received 24 May 2023; accepted 17 July 2023}

\abstract
{Luminous blue variable (LBV) stars represent a short-lived stage in the late evolution of the most massive stars. Highly unstable, LBVs exhibit dense stellar winds and episodic eruptions that produce complex circumstellar nebulae, whose study is crucial for properly constraining the impact of these sources at a Galactic scale from a structural, dynamical and chemical perspective.}
{We aim to investigate the molecular environment of \mbox{AFGL\,2298}, an obscured Galactic LBV which hosts a highly structured circumstellar environment with hints of multiple mass-loss events in the last few $10^4$ a.}
{We present spectral line observations of \mbox{AFGL\,2298} at 1 and 3 mm performed with the IRAM 30m radio telescope.}
{We report the detection of several carbon- and nitrogen- bearing species (CO, $^{13}$CO, C$^{18}$O, C$^{17}$O, HCO$^+$, HCN, HNC, H$^{13}$CO$^{+}$, CN, N$_2$H$^+$, and C$_2$H) in the surroundings of \mbox{AFGL\,2298}. We identified three velocity components that clearly stand out from the Galactic background. The morphology, kinematics, masses and isotopic ratios, together with a comparative study of the fractional abundances, lead us to suggest that two of these components (36 and 70 \kms) have a stellar origin. The other component (46 \kms) most likely traces swept-up interstellar material, probably harbouring also a photon-dominated region.}
{The first inventory of the circumstellar molecular gas around \mbox{AFGL\,2298} is provided. The results are compatible with the hypothesis of former mass-loss events, produced before the one that created the infrared nebula. There are chemical hints of the presence of ejected stellar material, and also swept up gas. These findings will help to better understand the mass-loss history of this class of evolved massive stars, which heavily influence the overall chemical evolution of the Galaxy.}


\titlerunning{Molecular gas in AFGL\,2298}
\authorrunning{Rizzo et al.}

   \maketitle
%
%-------------------------------------------------------------------

\section{Introduction}

The late evolutionary stages of massive stars play a key role in the structural, dynamical and chemical evolution of the Galaxy, releasing vast amounts of matter, radiative and mechanical energy into their surroundings \citepads{2012ARA&A..50..107L}. In this context, the luminous blue variable phase (hereafter LBV), a brief interlude of just a few \mbox{10$^4$\,a}, is of particular interest as it is responsible for a substantial fraction of the total mass loss. LBV stars are intrinsically variable and highly unstable objects that expel several solar masses of CNO processed material through steady, dense winds (typically from several $10^{-6}$ to few \mbox{$10^{-4}$\msun\,a$^{-1}$}) and episodic outbursts that rip off the outermost stellar layers in timescales as short as a few years \citepads{1994PASP..106.1025H}. All these physical processes create large and heterogeneous circumstellar nebulae of dust and gas, which are usually N--rich and C-- and O--deficient \citepads{2001ApJ...551..764L}. The study of these nebulae, therefore, is a potentially useful tool to reconstruct the mass-loss history of the central star.

Once identified as excellent laboratories to investigate the interplay between evolved massive stars and the interstellar medium (ISM), LBV nebulae have been the target of numerous observational campaigns at infrared and radio continuum wavelengths, tracing their dust and ionized gas content \citepads{1999LNP...523..381W, 2002MNRAS.330...63D, 2005A&A...437L...1U, 2009ApJ...694..697U, 2010ApJ...718.1036U}. These efforts have provided valuable insights about energetic budgets, mass-loss rates, nebular composition and dynamics. However, many questions remain unanswered: the phenomenology that triggers the eruptions is not totally understood, and the mechanisms that shape the expelled material (e.g. interaction with a binary companion, nearly critical rotation, magnetic fields) are yet to be fully established. The scarce number of LBVs, with $\sim$60 members identified in the Galaxy \citepads[of which only $\sim$20 are genuine LBVs, i.e., sources for which S~Dor variability has been confirmed][]{2018RNAAS...2..121R}, complicates even more any general conclusions.

In the past two decades, molecular spectroscopy has opened a new window to learn about LBVs and their mass loss processes. Despite the harsh conditions to which the outskirts of these stars are exposed (strong and variable far UV radiation beyond the Lyman limit, high temperatures), substantial amounts of molecular gas have been detected around a handful of LBVs \citepads[see e.g.,][]{2008ApJ...681..355R,2012ApJ...749L...4L,2020MNRAS.499.5269G}. This molecular component is believed to be a tracer of eruptive events, either being ISM material compressed by the expanding shock, or molecular gas formed \textit{in situ} from the CNO-processed ejecta. In both scenarios, the distribution, kinematics, and chemistry of the molecular gas shed light on key evolutionary aspects of the central star. For instance, slowly expanding, equatorial molecular rings of CO have been reported around \object{[GKF2010] MN101} \citep{2019MNRAS.482.1651B} and \object{AG Car} \citepads{2021MNRAS.500.5500B}, displaying low [$^{12}$CO/$^{13}$CO] isotopic ratios, a clear indicator of CNO-processing. These equatorial structures trace strong deviations from spherical symmetry, either due to binary interactions, or nearly critical rotation of the central star, which is the case for several LBV stars \citepads{2009ApJ...705L..25G,2011ApJ...736...46G}. On the other hand, more symmetric arrangements are found in other LBV sources (e.g., \object{G79.29+0.46}, \citealt{2008ApJ...681..355R, 2014MNRAS.440.1391A}, and \object{[GKF2010] MN48}, Bordiu et al., \textit{in prep}), in the form of shells which match the distribution of circumstellar dust; these structures are probably the result of the accumulation of ambient gas after an isotropic wind and/or mass eruption event.

The molecular inventory of LBV nebulae is however scarce, in most cases limited to CO and its main isotopologues, with the notable exceptions of \object{G79.29+0.46}, where also NH$_3$ and C$_3$H$_2$ have been detected \citep{2014A&A...564A..21R,2014ApJ...784L..21P}; and \object{$\eta$~Car}, that hosts a rich molecular chemistry, with confirmed detection of multiple C-, N- and O- bearing species (\citealt{2012ApJ...749L...4L,2019MNRAS.490.1570B,2020MNRAS.499.5269G,2020ApJ...892L..23M}) and, more recently, of Si-bearing molecules \citep{2022ApJ...939L..30B}.

\mbox{AFGL\,2298} (=IRAS 18576+0341) is a particularly good candidate to host important amounts of molecular gas. This blue supergiant was first proposed as a candidate member of the LBV class by \citet{2001ApJ...548.1020U}, owing to the discovery of an extended dusty shell of $\sim$7~arcsec through mid-IR observations, with morphological features compatible with an equatorially enhanced mass loss episode. Its LBV status was further reinforced \citepads{2002A&A...382.1005P} and eventually confirmed \citepads{2003A&A...403..653C} on the basis of multi-epoch spectroscopic and photometric observations, which revealed spectral features typical of LBV spectra and hints of S-Dor variability. \citetads{2003A&A...403..653C} determined $T_\mathrm{eff}$ variations between 12.5 and 15 kK in the period 2001--2002, with mass-loss rates ranging from $5\times10^{-5}$ to \mbox{$1.2\times10^{-4}$\,M$_\odot$\,a$^{-1}$}, at a roughly constant bolometric luminosity of $\log(L/L_\odot)$ = 6.2. These values were in close agreement with independent estimates by \citet{2005A&A...437L...1U} from multi-band VLA observations. Later, \citetads{2009A&A...507.1555C} combined archival observations with the results of a long-term monitoring campaign (2001--2008) and reported for the first time significant bolometric variability, interpreted as the result of an outburst instead of a typical S-Dor excursion. High resolution VLA and VLT observations \citepads{2010ApJ...721.1404B} provided an unprecedented view of the nebula around \mbox{AFGL\,2298} and revealed a rather homogeneous dusty shell coexisting with a strongly asymmetric ionized envelope. In the same study, it is even more interesting the discovery of polycyclic aromatic hydrocarbons (PAHs), also asymmetric but with a morphological distribution totally different with respect to the ionized gas. Despite this finding about the PAHs, the molecular content of \mbox{AFGL\,2298} remained unexplored.

In this paper, we present a comprehensive study of the molecular environment of \mbox{AFGL\,2298} at millimetric wavelengths. In Sect.~ \ref{sec:obs} we describe the observations; the main results are outlined in Sect.~\ref{sec:results}; in Sect.~\ref{sec:discussion} we place our findings in the wider context of LBV circumstellar chemistry, derive the physical parameters of the gas and the abundances of the observed species; and finally, in Sect.~\ref{sec:conclusions} we present the conclusions of this work and lay out the steps for follow-up studies.
   
%--------------------------------------------------------------------
\section{Observations} \label{sec:obs}

\mbox{AFGL\,2298} was observed with the IRAM-30\,m radio telescope of the Institut de RadioAstronomie Millim\'etrique in Granada (Spain), as part of project P043-17 (P.I: C. Bordiu). Observations took place on the nights of 2017 July 23 and 25 under good summer conditions ($\tau_\mathrm{225} \sim 0.2$). We used the EMIR multi-band receiver at its 3 and 1~mm bands (E090 and E230, respectively).

We selected the FTS backend, which provided an instantaneous bandwidth of 4 GHz per polarization, with an approximate channel width of $\sim$0.25 and $\sim$0.5 \kms\ at 3 and 1 mm respectively. The main goal was to simultaneously map the distribution of the $J=1\rightarrow0$ and $J=2\rightarrow1$ transitions of CO and its most abundant isotopologues. Additionally, we profited from the large bandwidth of the FTS backend in different spectral setups to cover all the rotational transitions indicated in Table~\ref{freqs}.

%--------------------------------------------------- Molecules observed
\begin{table}
  \caption[]{Rotational transitions of the molecules observed.}
  \label{freqs}
  \centering
  \begin{tabular}{llccc}
  \hline\hline
    \noalign{\smallskip}
    Molecule & transition & frequency & $E_\mathrm{up}$ & HPBW \\
             &            & GHz       & K                & arcsec \\
    \noalign{\smallskip}
    \hline
    \noalign{\smallskip}
    \multicolumn{5}{c}{CO and isotopologues} \\
    \noalign{\smallskip}
    CO        & $J=1\rightarrow0$ & 115.271202 &  5.5 & 21.3 \\
              & $J=2\rightarrow1$ & 230.538000 & 16.6 & 10.7 \\
    \noalign{\smallskip}
    $^{13}$CO & $J=1\rightarrow0$ & 110.201354 &  5.3 & 22.3 \\
              & $J=2\rightarrow1$ & 220.398684 & 15.9 & 11.2 \\
    \noalign{\smallskip}
    C$^{18}$O & $J=1\rightarrow0$ & 109.782173 &  5.3 & 22.4 \\
              & $J=2\rightarrow1$ & 219.560354 & 15.8 & 11.2 \\
    \noalign{\smallskip}
    C$^{17}$O & $J=1\rightarrow0$ & 112.358982 &  5.4 & 21.9 \\
              & $J=2\rightarrow1$ & 224.714187 & 16.2 & 10.9 \\
    \noalign{\smallskip}
    \multicolumn{5}{c}{Other molecules} \\
    \noalign{\smallskip}
    HCO$^+$        & $J=1\rightarrow0$ &  89.188525 & 4.3 & 27.6 \\
    HCN            & $J=1\rightarrow0$ &  88.631848 & 4.3 & 27.8 \\
    HNC            & $J=1\rightarrow0$ &  90.663568 & 4.4 & 27.1 \\
    H$^{13}$CO$^+$ & $J=1\rightarrow0$ &  86.754288 & 4.2 & 28.4 \\
    CN             & $N=1\rightarrow0$ & 113.191278 & 5.4 & 21.7 \\
                   & {\small $J=1/2\rightarrow1/2$} \\
    \noalign{\smallskip}
                   & $N=1\rightarrow0$ & 113.490970 & 5.4 & 21.7 \\
                   & {\small $J=3/2\rightarrow1/2$} \\
    \noalign{\smallskip}
    N$_2$H$^+$     & $J=1\rightarrow0$ &  93.173770 & 4.5 & 26.4 \\
    C$_2$H         & $N=1\rightarrow0$ &  87.316925 & 4.2 & 28.2 \\
                   & {\small $J=3/2\rightarrow1/2$} \\
    \noalign{\smallskip}
                   & $N=1\rightarrow0$ &  87.402004 & 4.2 & 28.1 \\
                   & {\small $J=1/2\rightarrow1/2$} \\
    \noalign{\smallskip}
    \hline

  \end{tabular}
  \tablefoot {Angular resolutions (HPBW) computed as 2460/frequency, according to the IRAM web pages (\url{https://publicwiki.iram.es/Iram30mEfficiencies}). As C$^{17}$O, HCN, CN, N$_2$H$^+$ and C$_2$H have hyperfine splitting, the quoted frequencies correspond to the most intense hyperfine component.}
\end{table}
%-------------------------------------------------------------------------------

The observing strategy involved on-the-fly (OTF) maps in position-switching mode, covering a square region of $1.5\times1.5$ arcmin around \mbox{AFGL\,2298}. Mapping was done in zigzag along two orthogonal directions, with a speed of 3 arcsec per second, and using a Nyquist-compliant subscan spacing of 4.3 arcsec. The reference position was located $\sim$20 arcmin away from \mbox{AFGL\,2298}, to minimise contamination from background or foreground diffuse emission. The total scan time was $\sim$2 hours. In addition, two deep integrations of $\sim$45 min were performed towards the star position and a random control position, offset (40, $-30$) arcsec from the star. At the time of observation, no small planet or strong quasar was observable close to the target, so Saturn was used instead for initial pointing and focus. In any case, we used the quasar 1749+046 --the closest in projection to \mbox{AFGL\,2298}-- for pointing in the vicinity of the source, reaching a accuracy better than 5 arcsec. G34.3+0.2 was used as line calibrator. Pointing was regularly checked during the observations, and calibrations were done between consecutive OTF runs.

The resulting calibrated spectra were then processed using the \texttt{GILDAS} software package. Spectra were first deplatformed (when needed) and baseline subtracted, and then combined to generate position-position-velocity data cubes for each transition and species. In this work, we adopt the following conventions: (1) spectra are presented in $T_\mathrm{A}^*$ scale, unless noted otherwise; (2) velocities refer to the local standard of rest (LSR) frame; and (3) positions are given as offsets from the J2000 coordinates of \mbox{AFGL\,2298}, which are ($\alpha$, $\delta$) = (19h00m10.9s, +3d45m47.2s).

%--------------------------------------------------------------------
\section{Results} \label{sec:results}
\subsection{Velocity ranges of interest}

\begin{figure}[hbt]
\centering
\includegraphics[width=0.95\columnwidth]{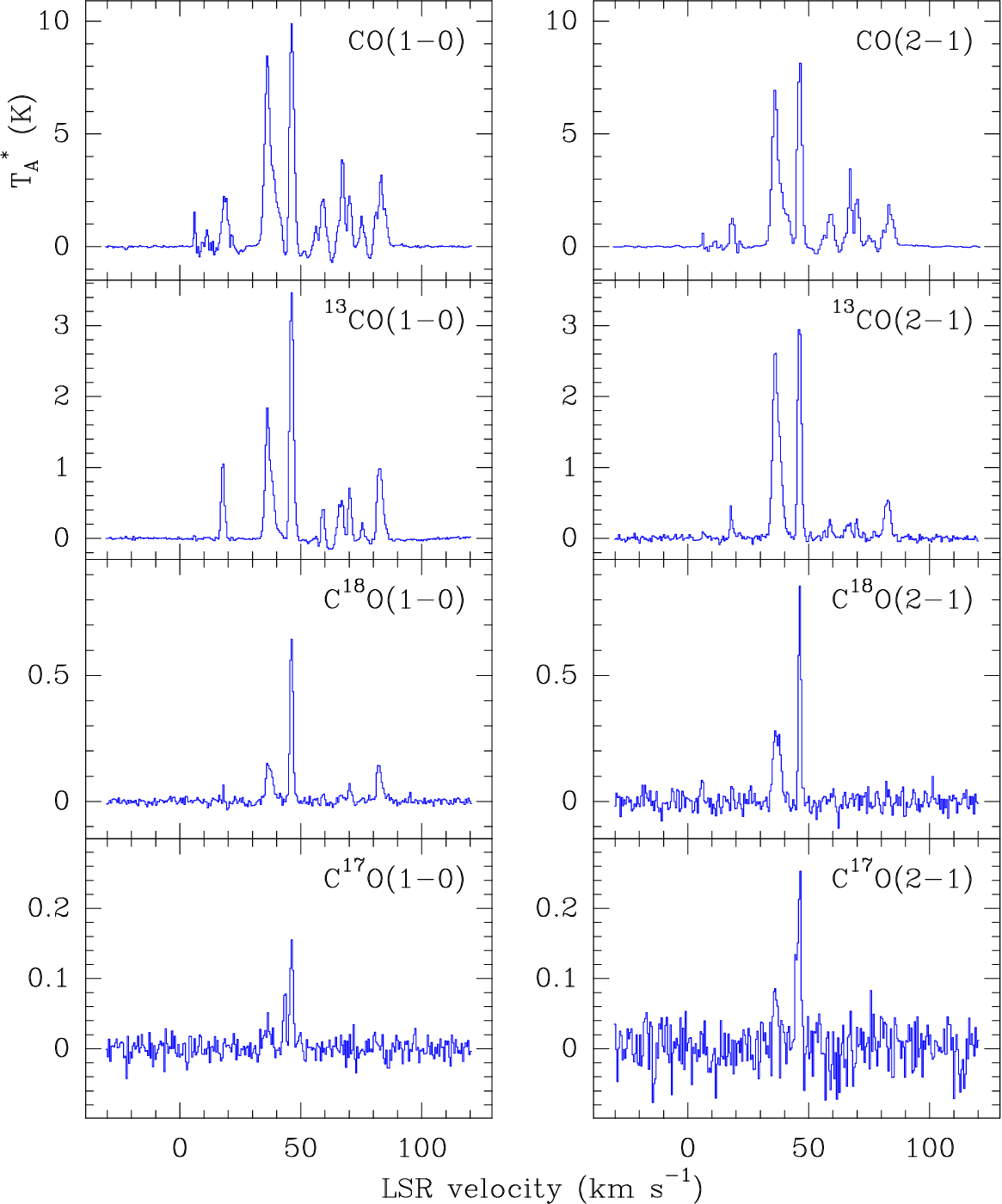}
\caption{CO and isotopologues in direction to \object{AFGL\,2298}. The species and their corresponding rotational lines are indicated in the upper-left corner of each spectrum. All spectra are displayed at the same velocity range. To ease a comparison, the $J=1\rightarrow0$ (left column) and the $2\rightarrow1$ (right column) lines are plotted on the same intensity scale. The most intense components, roughly at 38 and 46\kms\ (components A and B, see text) are the only ones present in all the lines. The C$^{17}$O lines have hyperfine structure, which is noted in the splitting at component B.}
\label{co}
\end{figure}

The $J=1\rightarrow0$ and $2\rightarrow1$ lines of CO and its most abundant isotopologues ($^{13}$CO, C$^{18}$O, and C$^{17}$O) towards the star position are depicted in Fig.~\ref{co}. Due to the location in the first quadrant of the galactic plane, the CO emission is complex and spread over a wide range of positive velocities. Being the tangent point velocity around 90\kms\ \citepads{1993A&A...275...67B, 2014ApJ...783..130R} no forbidden velocities are found.

As expected, all the velocity components fade or disappear when going from high to low abundance isotopologues (from top to bottom in Fig.~\ref{co}). The observed $J=2\rightarrow1$ to $1\rightarrow0$ line ratios are mostly lower than 1, as usual in cold galactic clouds.

The two most intense components at 32--38 \kms\ (hereafter component A) and at 44--50\kms\ (hereafter component B) depart from this trend and display a $J=2\rightarrow1$ intensity comparable or higher than their corresponding $J=1\rightarrow0$ one. These two components are also those which remain in the C$^{18}$O $J=2\rightarrow1$ and  C$^{17}$O lines. Component A is less intense and broader than component B, with some asymmetry and a ``shoulder'' at the most positive velocities ($47 - 50$\kms). Component B is nearly Gaussian and considerably narrow.

\begin{figure}
\centering
\includegraphics[width=0.95\columnwidth]{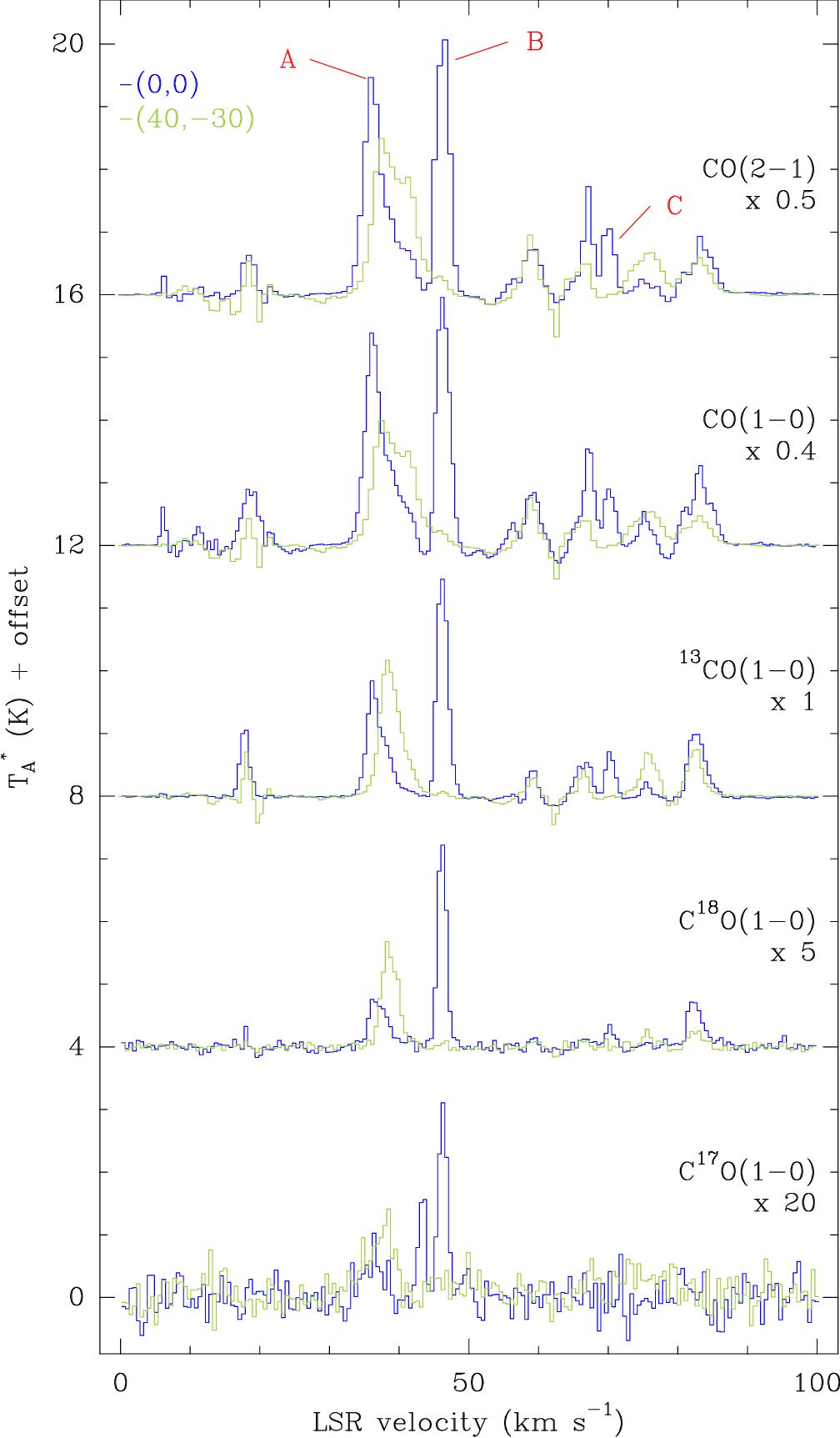}
\caption{Comparison of spectra observed at the star position and a control position outside the infrared nebula. The star position is plotted in blue, while the control position is in green. Different velocity components considered in this article are marked in red. The most remarkable features are the difference in the peak velocities in component A, and the lack of emission at the control position from components B and C.}
\label{onoff}
\end{figure}

The Fig.~\ref{onoff} displays all the CO isotopologues towards \mbox{AFGL\,2298} and the control position. As a first approach, their mutual comparison would help to disentangle the velocity components probably related to the target star. Component A is broadly present in the two positions. At the star position, however, it peaks systematically at a lower velocity with respect to the control position. The difference, noted only in this component and across all the isotopologues, may be ascribed to the presence of an additional feature related only to the star. On the contrary, component B, much narrower, is evident only in the pointing toward the star. A third component (approximately at 70\kms) is solely detected at the star position, and totally absent at the control position. We hereafter refer to this feature as component C.

Components B and C are clearly different in the two positions, which points to a possible relationship with the star. Although component A is more ubiquitous, the $J=2\rightarrow1$ to $1\rightarrow0$ line ratio is rather high and the line has some velocity structure; therefore, part of this extended component may also be affected by some interaction with the nebula and is not fully discarded. In the following sections, we concentrate the study on the three components identified. 

%  Figure 1 (co) --------------------------------------------------------
%

%  Figure 2 (onoff) --------------------------------------------------------

\subsection{Distribution of the CO gas}
To get more precise insights about the spatial distribution of the three components, we integrated the lines over their velocity extension. The results for the four $J=2\rightarrow1$ lines of the CO isotopologues are depicted in Fig.~\ref{all_comps}. The corresponding $J=1\rightarrow0$ lines display similar characteristics, although with poorer angular resolution.

Component A is quite extended along the whole region, with increasing intensity toward the northeast. Overall, it looks like a typical interstellar galactic cloud, without any evident features associated with the target star. In more detailed channel maps, however, a relative maximum of emission, at 10--15 arcsec from the centre, is detached from the general emission. This feature is consistent with the velocity difference noted already in the previous subsection.

Contrarily, component B has a certain symmetry with respect to the star position, although the peak is not coincident with the star. Overall, the component depicts an arc-like feature around the infrared nebula, extending approximately from northeast to south.

Component C is round and compact, and it is clearly centred at $\sim$15 arcsec toward the southwest of the hot star. Despite its small angular scale, this component is not a point source, which is evidenced after a comparison of the 50\% contour and the half power beam width.

Finally, it is worth noting that the intensity distribution is slightly different among the isotopologues, so that the kinematic components are stratified. This is particularly evident for component B, when comparing the positions of the relative maxima in Fig.~\ref{all_comps}: $^{13}$CO, C$^{18}$O and C$^{17}$O are located closer to the star, within the inner rim of the arc-like feature defined by CO, and also extending over a more limited angle. Besides, the relative maxima of C$^{18}$O and C$^{17}$O appear anti-correlated. Such a spatial distribution evidences a chemical stratification toward the northwest, which turns out to be one of the ``preferential'' directions of \mbox{AFGL\,2298} (see Sect.~\ref{subsec:relationship} below).

\subsection{Other molecules detected}
As mentioned in Sect.~\ref{sec:obs}, the large bandwidth of EMIR allowed us to observe some of the most simple molecules apart from the CO isotopologues (Table \ref{freqs}). We have made deep integrations towards the star position, and also on-the-fly maps covering roughly 90 arcsec around \mbox{AFGL\,2298}.

\begin{figure*}
\sidecaption
%\centering
\includegraphics[width=12cm]{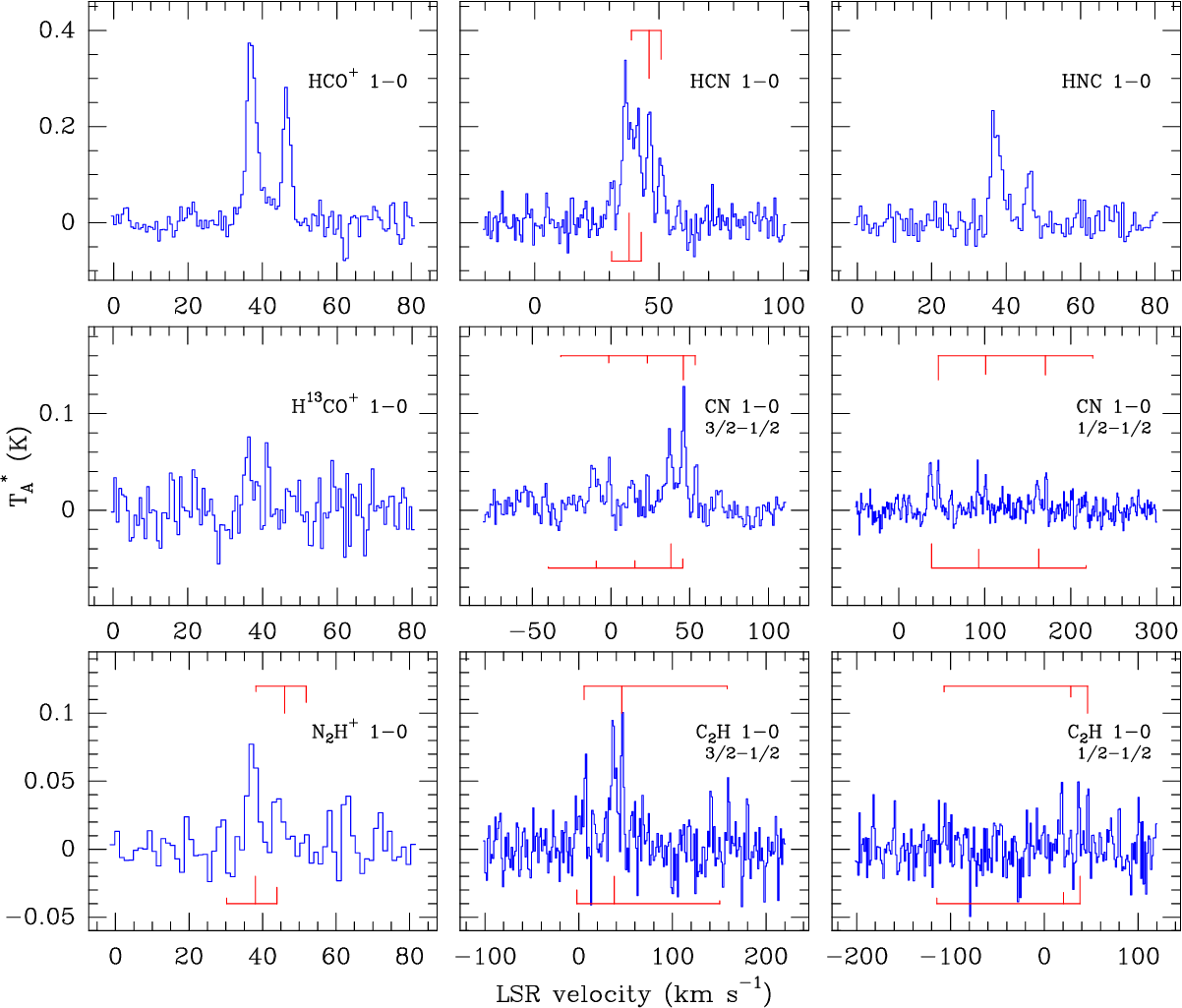}
\caption{Other molecules detected in the direction to \object{AFGL\,2298}. Molecule and transition are indicated inside each spectrum. Note that different velocity ranges are depicted, and that the temperature scale is kept the same for each row. In the case of molecules having hyperfine splitting, the position and relative intensities are indicated by the red lines below (above) the spectrum, for velocities of 36 (46) \kms.
 }
\label{other_molec}
\end{figure*}

Figure \ref{other_molec} depicts the resulting spectra of these molecules towards the star position. For those molecules with hyperfine splitting, red lines indicates position and relative sizes of their components. The most intense lines are HCO$^+$ and HCN, two of the most abundant molecules in all environments besides H$_2$ and CO.

Both components A and B are present in all the spectra, except N$_2$H$^+$, where component B is not detected. Component C, on the other hand, is not detected in any of the nine molecular lines, probably due to insufficient sensitivity. Like in the case of CO isotopologues, component A seems wider than component B. Outstandingly, the relative intensities between these two components is different for each molecule: component A is more intense in HCO$^+$, HCN, HNC and N$_2$H$^+$, and similar or even less intense in the others (see the case of CN in the central panel).

The maps are rather noisy and with poor angular resolution, but sufficient to give a first idea about the distribution of the seven molecules, especially for the component A. Detailed maps, integrated in the velocity ranges of the three components, are provided in the Appendix. Component A is well traced by several molecules, and depicts different distributions, preferentially to the north and the east of the region mapped. Component B is clearly noted in HCO$^+$, HCN, HNC, and the most intense CN and C$_2$H transitions ($N=1\rightarrow 0; J=3/2\rightarrow 3/2$ in both cases), with a morphology roughly similar to the CO isotopologues. The non-detection of H$^{13}$CO$^+$ and N$_2$H$^+$ is consistent with the results obtained in the deep on-source integrations, described above. Anyway, it should be stressed that the angular resolution and the noise level of the maps are not sufficient to firmly establish morphological differences among the observed molecules.

\subsection{Kinematics}
\begin{figure}
\centering
\includegraphics[width=0.95\columnwidth]{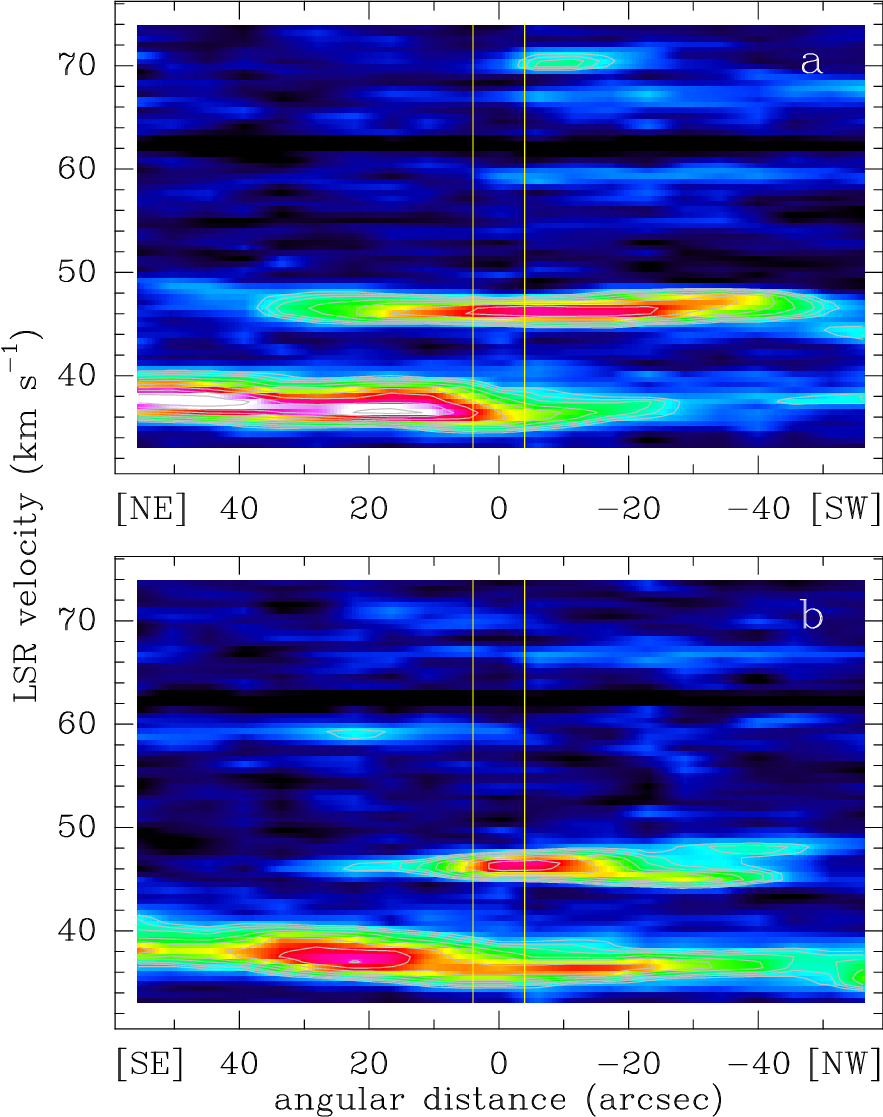}
\caption{Position-velocity plots of the $^{13}$CO $J=2\rightarrow1$ line. Slices are traced (a) from northeast $(+40^{\prime\prime}, +40^{\prime\prime})$ to southwest $(-40^{\prime\prime}, -40^{\prime\prime}$); and (b) from southeast $(+40^{\prime\prime}, -40^{\prime\prime})$ to northwest $(-40^{\prime\prime}, +40^{\prime\prime}$). Contour levels are 0.6, 0.75, 0.9, 1.2, 1.5, 2.0, 2.5, 3.0, and 3.5~K\,\kms. Vertical yellow lines mark the approximate extension of the infrared dust nebula. In plot (a) the three components are clearly noted, being component C and part of the component A just outside the nebula. In plot (b), perpendicular to the previous one, component C is not present and component B is concentrated on the nebula.
}
\label{pv}
\end{figure}

The kinematics of the velocity components may, in principle, be inferred from position-velocity (PV) plots. In this case, the most adequate molecule to study the structure in velocity is $^{13}$CO because it is less affected by galactic contamination than CO and, at the same time, with sufficient intensity to clearly distinguish the three components. Figure \ref{pv} displays two $^{13}$CO $J=2\rightarrow1$ PV plots in orthogonal directions: panel (a) from northeast to southwest includes both the peak of component A and the whole component C, and panel (b) which crosses the component B. 

In the first PV plot, component C again appears as a compact and weak feature, located right outside the infrared nebula (roughly traced by the vertical yellow lines). It is also notable that, in component A, the emission from 38 to 40\kms\ is concentrated at the northeast, opposite to component C. In the second PV plot it is possible to note that, as inferred also by the previous figures, component B seems more concentrated towards the centre of the field.

Despite the interesting nature of these results, we are far from disentangling the dynamics of the molecular gas due to an evident lack of angular resolution. This source is indeed an excellent target to be observed by interferometry due to its richness in molecular species and also because it hosts several molecular lines with relatively high intensities.

%  Figure 3 (all_comps)  ------------------------------------------------
\begin{figure*}
\centering
\includegraphics[width=0.95\textwidth]{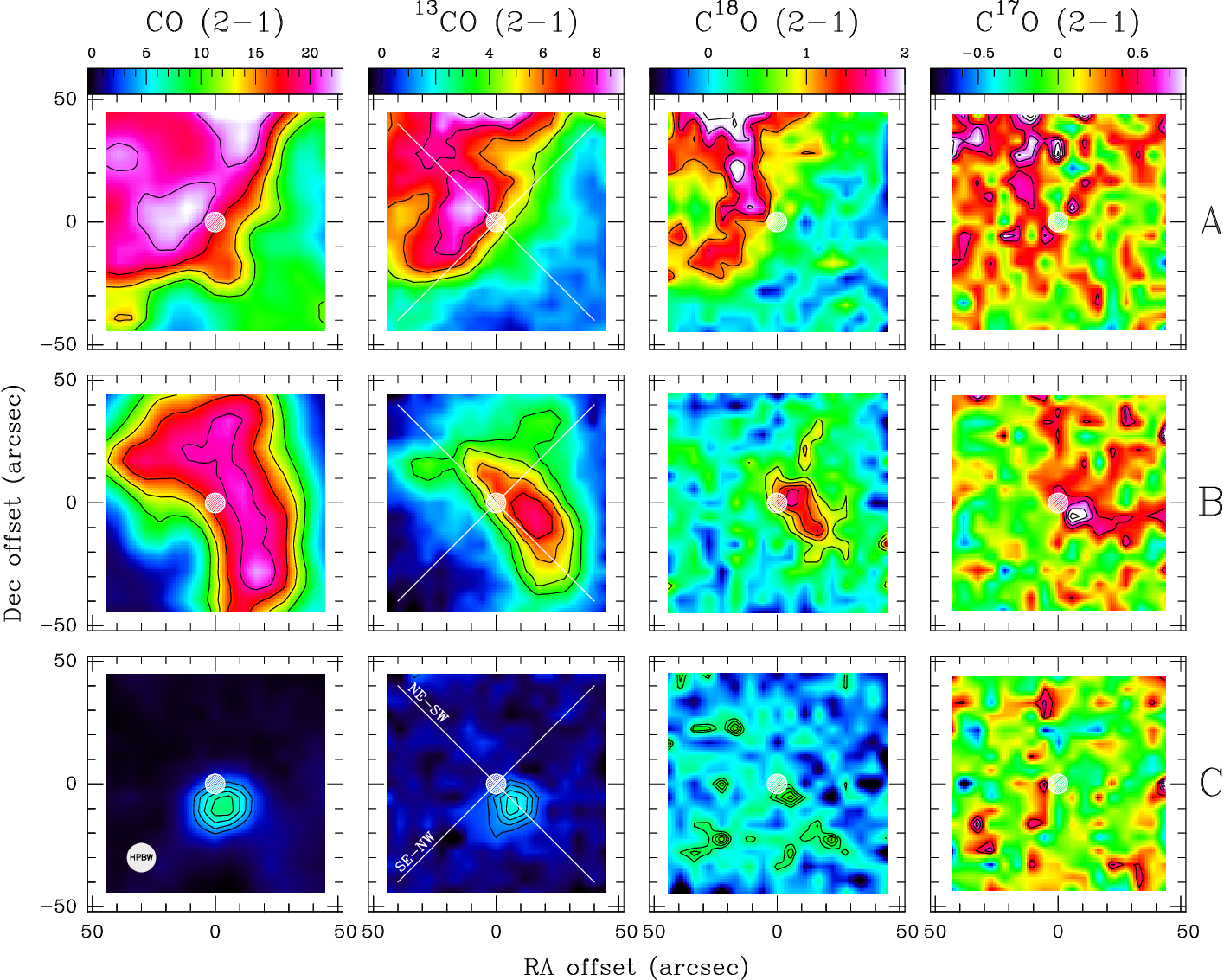}
\caption{Overview of the intensity distribution of the components analysed. Top, middle and bottom rows correspond to components A, B and C, respectively. Velocity ranges of integration are (35, 38), (45, 48), and (69, 72)\kms. Spectral lines are indicated above the top maps. Equatorial coordinates are relative to the star position. The colour scale is the same for all three components of a given molecule, and is indicated on the top panels, in units of \mbox{K\,\kms}. Contours are 45, 60, 75 and 90\% of the peak values. The white circle sketches the position and size of the infrared nebula. Half power beam width is indicated in the bottom-left corner of the CO(2-1) map of component C. The white lines plotted onto the $^{13}$CO maps indicate the direction of the position-velocity plots of Fig.~\ref{pv}.
}
\label{all_comps}
\end{figure*}
%

%  Figure 4 (other_molec) --------------------------------------------------------
%

%  Figure 5 (pv) --------------------------------------------------------
%

%  Figure 6 (vla) ------------------------------------------------
\begin{figure*}
\sidecaption
%\centering
\includegraphics[width=12cm]{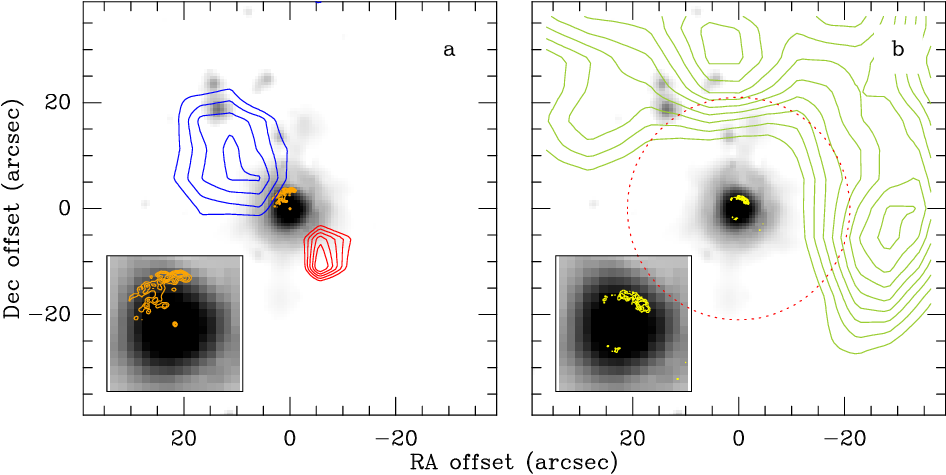}
\caption{
Relationship of the kinematic molecular components with the warm dust and the ionized gas. Greyscale corresponds to the Spitzer/IRAC band 3 (5.7~$\mu$m) and clearly indicates the extension of the infrared nebula. 
(a) Blue and red contours correspond to components A and C in the $^{13}$CO $J=2\rightarrow1$ line. Level are from 70 to 95~\% of the peak value (5.68 and 1.76~K\kms, respectively), in steps of 5~\%. Orange contours correspond to 6-cm continuum emission \citepads{2010ApJ...721.1404B}, from 0.6 to 1~mJy, in steps of 0.1~mJy. 
(b) Green contours depict component B in the CO $J=2\rightarrow1$ line, from 2 to 9~K\kms, in steps of 0.5~K\kms. Yellow contours correspond to the continuum-subtracted PAH emission at 11.26~$\mu$m \citepads{2010ApJ...721.1404B}, from 0.7 to 1.3~mJy, in steps of 0.2~mJy. The red circle indicates the deconvolved size of the radio recombination line emission reported by \citetads{2015ApJS..221...26A}. The insets in both panels contain the same greyscale and contours, but zoomed to the central 10~arcsec to ease the visualization. 
}
\label{vla}
\end{figure*}
%

%  Figure 7 (regions) ------------------------------------------------
\begin{figure*}
%\sidecaption
\centering
\includegraphics[width=0.9\textwidth]{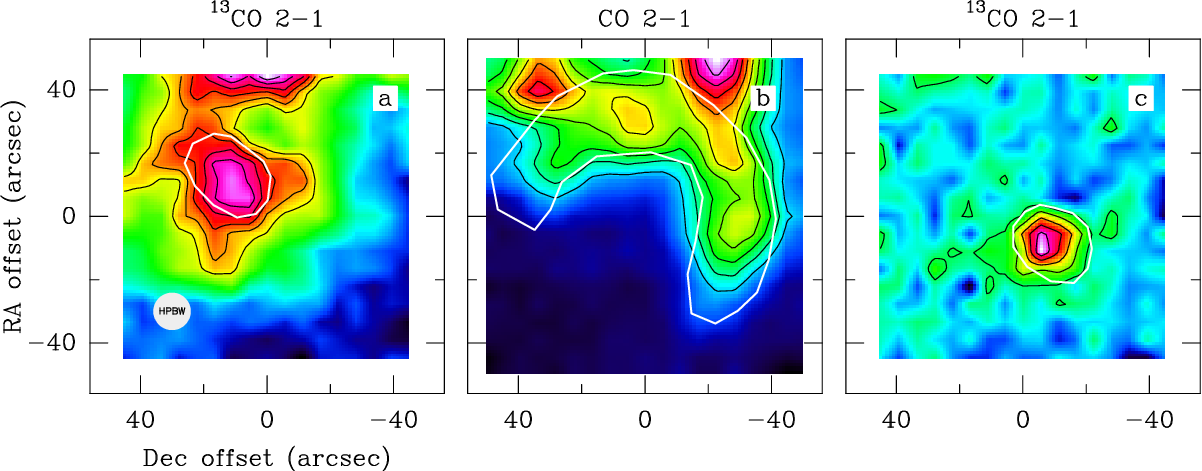}
\caption{
Regions defined for the column density estimates of the three velocity components, sketched as polygons in white. Lines, velocity intervals of integration, first contours and spacing are: (a) $^{13}$CO $J=2\rightarrow1$, 35.4 to 37.0\kms, 3~K\kms, and 0.4~K\kms; (b) CO $J=2\rightarrow1$, 
47.6 to 49.1\kms, 1.4~K\kms, and 0.7~K\kms; (c) $^{13}$CO $J=2\rightarrow1$, 
69.6 to 70.7\kms, 0.3~K\kms, and 0.3~K\kms.
Half power beam width is shown near the bottom-left corner of figure a.
}
\label{regions}
\end{figure*}

%-----------------------------------------------------------------
\section{Discussion} \label{sec:discussion}

%-------------------------------------------------------------------------------
\begin{table*}
\caption{Distances.}
\centering
\begin{tabular}{lllll}
  \hline\hline
  \noalign{\smallskip}

   Data source              & method       & $d$ (kpc) & error (kpc) & comment                                 \\
    \noalign{\smallskip}
    \hline
    \noalign{\smallskip}
Ueta et al.~(2001)                & model                   & 10.0     & 3     & Distance range necessary to fit IR data.       \\
Ram\'irez Alegr\'ia et al.~(2018) & spectro-photometry      & 9.6      & 0.4   & Based on a stellar group at similar distances. \\
Ram\'irez Alegr\'ia et al.~(2018) & galactic rotation model & 9.3      & 1.9   & See note 1.                   \\
Anderson et al.~(2015)            & galactic rotation model & 10.7     & 0.9   & See note 2.       \\
This work                         & galactic rotation model & 10.6     & 0.4   & Including full range of components A and B.    \\
\noalign{\smallskip}
Adopted value                     & weighted average        & {\bf 10.1}     & {\bf 0.6}   & \\
\noalign{\smallskip}
    \hline
\end{tabular}
\tablefoot{
(1) The distance uncertainty quoted results after considering a velocity uncertainty of 21\kms, which is the standard deviation of the radial velocities of the members of Masgomas-6b. (2) Uncertainty after assuming 15\kms\ of line velocity uncertainty, which is half of the HPBW of the radio recombination lines.
}
\label{distances}
\end{table*}
%-------------------------------------------------------------------------------

%--------------------------------------------------- Column densities
\begin{table*}[b]
  \caption[]{Modelling and parameters}
  \label{coldens}
  \centering
  
(a) \texttt{RADEX}/\texttt{ndradex} results\\{\smallskip}

  \begin{tabular}{ccrrrcrcc}
  \hline\hline
    \noalign{\smallskip}{\smallskip}
Comp.	&	Vel.\ range	&	$n$(H$_2$)	&	$T_{\mathrm k}$	&	mass	&	$N$(CO)	&	$N$($^{13}$CO)	&	$N$(C$^{18}$O)	&	CO/$^{13}$CO	\\
	&	\kms	&	$10^4$ cm$^{-3}$	&	K	&	M$_\odot$	&	$10^{16}$ cm$^{-2}$	&	$10^{15}$ cm$^{-2}$	&	$10^{14}$ cm$^{-2}$	&		\\
    \noalign{\smallskip}
    \hline
    \noalign{\smallskip}
A	&	$(35.4, 37.0)$	&	$0.5 \pm 0.1$	&	$24 \pm 4$	&	$1.0 \pm 0.2$	&	$0.30 \pm 0.06$	&	$1.39 \pm 0.06$	&	$5.6 \pm 0.2$	&	2.2	\\
B	&	$(47.6, 49.1)$	&	$0.6 \pm 0.2$	&	$19 \pm 5$	&	$78 \pm 22$	&	$4.7 \pm 1.3$	&	$4.9 \pm 0.5$	&	$3.9 \pm 0.3$	&	9.6	\\
C	&	$(69.4, 71.0)$	&	$2.1 \pm 0.9$	&	$18 \pm 5$	&	$1.4 \pm 0.2$	&	$0.47 \pm 0.05$	&	$1.09 \pm 0.07$	&	$1.20 \pm 0.02$	&	4.3	\\
    \noalign{\smallskip}
    \hline
  \end{tabular}
  %\tablefoot {blah.}
\vspace{3mm}
\\

(b) \texttt{MADCUBA} results\\{\smallskip}

  \begin{tabular}{ccrrrrccc}
  \hline\hline
    \noalign{\smallskip}{\smallskip}
Comp.	&	$N$(C$^{17}$O)	&	$N$(HCO$^+$)	&	$N$(HCN)	&	$N$(HNC)	&	$N$(H$^{13}$CO$^+$)	&	$N$(CN)	&	$N$(N$_2$H$^+$)	&	$N$(C$_2$H)	\\
	&	$10^{14}$ cm$^{-2}$	&	$10^{12}$ cm$^{-2}$	&	$10^{12}$ cm$^{-2}$	&	$10^{12}$ cm$^{-2}$	&	$10^{11}$ cm$^{-2}$	&	$10^{13}$ cm$^{-2}$	&	$10^{11}$ cm$^{-2}$	&	$10^{13}$ cm$^{-2}$	\\
    \noalign{\smallskip}
    \hline
    \noalign{\smallskip}
A	&	$1.9 \pm 0.5$	&	$2.7 \pm 0.3$	&	$8.2 \pm 0.7$	&	$2.2 \pm 0.2$	&	$<1.6$	&	$1.2 \pm 0.3$	&	$9 \pm 1$	&	$4.0 \pm 0.8$	\\
B	&	$0.9 \pm 0.3$	&	$0.9 \pm 0.2$	&	$1.74 \pm 0.07$	&	$0.94 \pm 0.1$	&	$<0.76$	&	$0.5 \pm 0.1$	&	$<1.2$	&	$1.1 \pm 0.3$	\\
C	&	$<0.42$	&	$<0.12$	&	$<0.39$	&	$<0.16$	&	$<1.4$	&	$<0.15$	&	$<2.1$	&	$<0.80$	\\
    \noalign{\smallskip}
    \hline
  \end{tabular}

\end{table*}
%-------------------------------------------------------------------------------

%-------------------------------------------------------------  Abundances
\begin{table}
\caption{Average fractional abundances with respect to H$_2$.}
\centering
\begin{tabular}{lccc}
  \hline\hline
    \noalign{\smallskip}{\smallskip}
Species & A                 & B                  & C                 \\
    \noalign{\smallskip}
    \hline
    \noalign{\smallskip}
$^{13}$CO       & $3.5\times10^{-5}$            & $8.3\times10^{-6}$        & $1.9\times10^{-5}$ \\
C$^{18}$O       & $1.5\times10^{-5}$            & $6.6\times10^{-7}$        & $2.1\times10^{-6}$ \\
C$^{17}$O       & $5.1\times10^{-6}$            & $1.5\times10^{-7}$        & $<7.3\times10^{-7}$ \\
HCO$^+$         & $7.2\times10^{-8}$            & $1.5\times10^{-9}$        & $<2.1\times10^{-9}$ \\
HCN             & $2.2\times10^{-7}$            & $3.0\times10^{-9}$        & $<6.8\times10^{-9}$ \\
HNC             & $5.8\times10^{-8}$            & $1.6\times10^{-9}$        & $<2.8\times10^{-9}$ \\
H$^{13}$CO$^+$  & $<4.3\times10^{-9}$           & $<1.3\times10^{-10}$      & $<2.4\times10^{-9}$ \\
CN              & $3.2\times10^{-7}$            & $8.5\times10^{-9}$        & $<2.6\times10^{-8}$ \\
N$_2$H$^+$      & $2.4\times10^{-8}$            & $<2.0\times10^{-10}$      & $<3.7\times10^{-9}$ \\
C$_2$H          & $1.1\times10^{-6}$            & $1.7\times10^{-8}$        & $<1.4\times10^{-7}$ \\
    \noalign{\smallskip}
    \hline
\end{tabular}
\tablefoot{
\tablefoottext{*}{Assumed a [CO/H$_2$] ratio of $8\times10^{-5}$.}
}
\label{abundances}
\end{table}
%-------------------------------------------------------------------------------

\subsection{Distance to \mbox{AFGL\,2298}}
The first estimate of the distance to \mbox{AFGL\,2298}, $10\pm3$ kpc, arises from \citetads{2001ApJ...548.1020U}. This ample range is the necessary to fit the observed infrared features in the nebula, under the assumption of a LBV nature of the exciting star. Despite the large uncertainty, subsequent works continued assuming 10~kpc due to the impossibility of further estimates \citepads{2003A&A...403..653C, 2009A&A...507.1555C, 2010ApJ...721.1404B}.

The stellar light is highly absorbed by interstellar extinction and also by extinction from the surrounding nebula. At visual wavelengths, we searched in the literature and the catalogues using VizieR\footnote{\url{https://doi.org/10.26093/cds/vizier}}, and only found firm measurements from Pan-STARRS \citep[objID 112512850453746126;][]{2016arXiv161205560C}, just in the z and y bands, with apparent magnitudes of $20.4\pm 0.1$ and $17.5\pm 0.1$~mag, respectively. \mbox{AFGL\,2298} is not included in any release of the Gaia mission \citep{2016A&A...595A...2G, 2018A&A...616A...1G, 2023A&A...674A...1G}, so it is not possible to obtain any distance estimate by direct measurement of its parallax.

One of the indirect methods to assign a distance is by establishing a membership to a stellar cluster or association with known distance. By combining high resolution infrared spectroscopy and high sensitivity photometry from 2MASS and UKIDSS, \citetads{2018A&A...614A.116R} reported the existence of a stellar population --Masgomas-6b-- at a distance of $9.6\pm0.4$~kpc, which includes \mbox{AFGL\,2298} as a member.

Interestingly, \citetads{2018A&A...614A.116R} also reports 65\kms\ as the mean velocity of Masgomas-6b, although \mbox{AFGL\,2298} does not contribute to this value. The velocity of Masgomas-6b is also consistent with its quoted distance: according to the galactic rotation model of \citet{2014ApJ...783..130R}, the LSR velocity of Masgomas-6b corresponds to a far distance of $9.3^{+1.2}_{-1.9}$~kpc.

Another key result to better constrain the distance is the detection of radio recombination lines (RRLs) reported by \citet{2015ApJS..221...26A}, in their survey of WISE sources having infrared features consistent with {\sc Hii} regions\footnote{The catalogue presents a stack of RRLs from H87$\alpha$ to H93$\alpha$.}. We searched the catalogue using VizieR (catalogue \texttt{J/ApJS/221/26/table4}) and identified the source \mbox{G037.277-00.226} as the ionized counterpart of \mbox{AFGL\,2298}, with two LSR velocities reported: $41.8\pm0.2$ and $77.7\pm0.9$\kms. The first velocity lies in the middle of those corresponding to components A and B, while the second one is $\sim8$\kms\ apart from the component C. In the same archive, the quoted kinematic distances correspond to 41.8\kms, being  2.8 and 10.7~kpc the near and far distances, respectively. The distance ambiguity is solved in favour of the far distance to keep consistency with the electron temperature derived from their own continuum observations.

The kinematic distances associated with components A and B, in very good agreement with all the results previously discussed, reinforces the link of these components to our target star. The velocity range from 37.3 to 47.9\kms\ translates to a distance range 10.3 to 10.9~kpc using, again, the model of \citet{2014ApJ...783..130R}. Taken altogether, our CO data are therefore consistent with a kinematic distance of $10.6 \pm 0.4$~kpc.

To summarize, we compiled in Table \ref{distances} all the individual distance estimates, together with their uncertainties. The photometric distance from \citetads{2018A&A...614A.116R} allow us to definitely adopt the far kinematical distance, which reliably may be established as $10.1 \pm 0.6$~kpc, as indicated in the last row of the table.

\subsection{Molecular gas related to the star}
\label{subsec:relationship}

As explained in the above sections, the three components display a morphology with a peculiar location with respect to the star and the infrared nebula: components A and C are compact and close to the star, while component B is arc-like, with the hot star approximately at its centre. In addition, two of these components (A and B) have velocities compatible with the inferred distance to \mbox{AFGL\,2298}, and are the only ones that have emission in other molecules.

There are two preferential directions, which are the ones chosen for the PV plot displayed in Fig.~\ref{pv}. The first direction, approximately from northeast to southwest, includes the peak of components A and C, which are symmetrically located with respect to the star and the nebula. The second one, roughly perpendicular, is rather well aligned with the symmetry axis of the component B. Interestingly, \citetads{2010ApJ...721.1404B}, who observed radio continuum at 6 and 20~cm with VLA, and five infrared bands from 11.26 to \mbox{17.65\,$\mu$m} with VLT/VISIR, already identified this lack of symmetry. In the Fig.~\ref{vla} we compare the relative position of the circular nebula as seen by the band 3 of Spitzer/IRAC, our data, and the \citetads{2010ApJ...721.1404B}'s continuum VLA and VLT observations.

In the Fig.~\ref{vla}(a) we plot the $^{13}$CO $J=2\rightarrow1$ line as blue and red contours for components A and C, and also the most intense part of the continuum emission at 6~cm. As expected, the [Ne{\sc ii}] emission is well correlated to the radio continuum, and it is not shown here for clarity. The three features (radio continuum, and components A and C) are very well aligned in the northeast-southwest direction. Interestingly, the alignment may support the shock-ionization scenario proposed by \citet{2010ApJ...721.1404B}  as an explanation for the asymmetric 6-cm emission, indicating the presence of denser material in the NE-SW direction. 

In the Fig.~\ref{vla}(b) we show as green contours the CO $J=2\rightarrow1$ line emission representative of the component B (a single spectral channel at 48.4\kms), and the PAH emission at 11.26 $\mu$m as yellow contours. As shown in the inset, the most intense emission from PAH arises in the innermost region and is directed towards the northwest, in excellent agreement with the CO. PAH emission is also correlated with the dust emission at \mbox{17.65~$\mu$m}. As discussed by \citet{2010ApJ...721.1404B}, the observed asymmetry could be tracing significant anisotropies in the mass-loss event that produced the dusty nebula, in principle different from the one that triggered the shock-ionisation towards the northeast.

It is worth noting that the catalogue of \citetads{2015ApJS..221...26A} reports the existence of extended radio emission in the position of \mbox{AFGL\,2298}, having an angular size of 97~arcsec. After deconvolving this size by the beam of the GBT telescope at 9~GHz (87~arcsec), it is obtained an angular diameter of 42~arcsec. This feature, sketched by a dashed red circle in Fig.~\ref{vla}(b), is notoriously tracing the inner part of the component B.

If all these components are connected, the CO is tracing older mass-loss events and the emission associated with the infrared nebula are clearly younger. By assuming a characteristic velocity of 15\kms\ (approximately the geometric mean of the differences between A-B and B-C components), and a size of 0.6~pc (corresponding to 12 arcsec at 10~kpc), we obtain a characteristic dynamic time of \mbox{$3.8 \times 10^4$\,a}, compatible with the LBV phase. A crude estimate of the age of the nebula may be deduced from the width of the Br$\gamma$ line \citepads{2009A&A...507.1555C}, which is 70\kms, and a radius of 0.2~pc, corresponding to 4~arcsec at 10~kpc. The resulting dynamical time for the nebula results of about \mbox{2700\,a}. Therefore, the molecular gas reported here represents the relics of older mass-loss events which have been produced in the same direction as the newer ones, traced by the ionized gas inside the nebula.

By increasing the level of speculation, component B may constitute the remnants of an equatorially enhanced mass-loss event (a ring), while components A and C may represent polar mass ejections. The lack of symmetry in velocity (component B is not halfway between components A and C) may be explained by an non-homogeneous expansion; the receding part of the jet may have encountered less resistance to expansion because it may have reached the border of the natal molecular cloud, where the approaching part is still embedded. This picture is also consistent with the lack of space symmetry depicted by radio continuum and recombination lines \citepads{2005A&A...437L...1U, 2010ApJ...721.1404B, 2009A&A...507.1555C}.

\subsection{Radiative transfer modelling}

\subsubsection{CO, $^{13}$CO and C$^{18}$O} \label{ndradex}
In principle, the molecular gas studied here is not seriously affected by radiative excitation from \mbox{AFGL\,2298} or any other nearby exciting sources. Therefore, the gas should be primarily excited by collisions and close to local thermodynamic equilibrium (LTE). However, the time necessary to reach LTE in this environment is not well determined. Our data allow us not only to estimate global physical conditions of the gas, but also to explore whether the LTE assumption may be applicable to it.

To determine the physical parameters of the gas in the three velocity components, we employed the \texttt{ndradex} package\footnote{\url{https://github.com/astropenguin/ndradex}}, a Python wrapper of the state-of-the-art radiative transfer code \texttt{RADEX} \citep{2007A&A...468..627V}. Instead of assuming LTE, \texttt{RADEX} applies the escape probability formulation \citep{1960mes..book.....S} to solve the radiative transfer equation under non-LTE conditions, thus predicting molecular line intensities and level populations for a given set of physical conditions. \texttt{ndradex} allows us to explore large grids of models in a space parameter defined by the column density of the molecule(s) under analysis $N(X)$, the H$_2$ volume density $n(\mathrm{H}_2)$, and the kinetic temperature of the gas $T_\mathrm{k}$.

To compute the global physical parameters $n(\mathrm{H}_2)$ and $T_\mathrm{k}$, we relied on $^{12}$CO, $^{13}$CO, and C$^{18}$O. We took advantage of having both the $J=1\rightarrow0$ and $J=2\rightarrow1$ transitions measured for the three species, which allows us to simultaneously fit six lines, minimising the  degeneracy between $T_\mathrm{k}$ and $n(\mathrm{H}_2)$. Using the velocity ranges where the three components are seen more clearly, we defined a characteristic region for each component in its corresponding integrated intensity map (see Fig. \ref{regions}). Then, we averaged the spectra of each molecule in each region, significantly improving S/N. The resulting spectra were finally used for comparison with the synthetic spectra produced by \texttt{ndradex}. To find the best fitting model, i.e. the set of physical parameters that better reproduces the observed line intensities for all the six lines observed, we applied a reduced $\chi^2$ minimisation, defined as:

\begin{equation}
\chi^2=\sum_{i=1}^3 \sum_{J=1}^2\left(\frac{I_{i J}^{\mathrm{obs}}-I_{i J}^{\mathrm{mod}}}
{\Delta I_{i J}}\right)^2
\end{equation}

\noindent where $i$ stands for the different isotopologues and $J$ represents the transition $J\rightarrow J-1$. $I_{i J}^{\mathrm{obs}}$ and $I_{i J}^{\mathrm{mod}}$ are the line intensities in main beam temperature scale\footnote{Computed as $T_\mathrm{MB} = T_\mathrm{A}^*/\eta_\mathrm{eff}$, where $\eta_\mathrm{eff}$ is the main beam efficiency obtained from \url{https://publicwiki.iram.es/Iram30mEfficiencies}} ($\int {T_{\mathrm MB}\ dv}$) observed and predicted by \texttt{RADEX}, respectively. $\Delta I_{i J}$ is the uncertainty in $I_{i J}^{\mathrm{obs}}$. 

Since the spectra are averaged over regions that are substantially larger than the \mbox{IRAM-30\,m} beam, it was not necessary to convolve the data to a common angular resolution, nor to apply a correction by beam filling factor. Because we are providing mean values over each region, it is implicitly assumed that the emission from all the lines is co-spatial and shares the same kinematics. We have therefore adopted a common linewidth for all the species in each region, computed as the weighted average of the individual linewidths, and defining the weights as the inverse squared of the individual uncertainties.

We employed a two-step approach for the fitting. First, we built a coarse grid of $\sim$17000 models for each component, with $T_\mathrm{k}$ in the range 10--100 K, $N(\mathrm{CO})$, $N(\mathrm{^{13}CO})$ and $N(\mathrm{C^{18}O})$ in the range 10$^{13}$--10$^{19}$ cm$^{-2}$, and $n(\mathrm{H_2})$ in the range 10$^1$--10$^5$ cm$^{-3}$. Within this coarse grid, we found several solutions (i.e., relative minima of $\chi^2$) for all the three velocity components. In some of the solutions, we found that the excitation temperature ($T_\mathrm{ex}$) reaches values without physical meaning, i.e., negative or excessively high. Taking into account that there is not a significant source of radiative excitation (the gas is located far away of the star), the molecular gas modelled is primarily excited by collisions and $T_\mathrm{ex}$ should not be too different to the $T_\mathrm{k}$ used as input. In consequence, we adopt the criterion to exclude those solutions with $T_\mathrm{ex} > 1.5 T_\mathrm{k}$.

Once identified the regions of the parameter space that produced physically plausible models, we recomputed the grids around those regions using smaller sampling steps ($\Delta T_\mathrm{k}$ = 5 K, $\Delta n(\mathrm{H}_2)$ = 500 cm$^{-3}$, $\sim$2500 models per component). In all these models, we observed that $T_\mathrm{ex}$ of the $J=1\rightarrow0$ transition is always moderately higher than that of the $J=2\rightarrow1$ lines. This difference may indicate some small departure from LTE conditions.

For each component, we adopted the average and standard deviation of $T_\mathrm{k}$, $n(H_2)$ and $N(\mathrm{X})$ of the best models (i.e. around the absolute $\chi^2$ minimum) as the final values and their associated uncertainties. 

As we noted in Sect. \ref{sec:results}, the detached emission feature that defines component A (i.e., the region A of Fig.~\ref{regions}) is embedded in an extended emission plateau, probably unrelated to the star. This plateau may bias the column densities toward higher values. To deal with this bias, we subtracted the contribution of the plateau by a crude estimate of its intensity over several regions right outside region A. Table \ref{coldens}(a) shows the values adopted as the best fitting cases with physical meaning.

The densities of the three components are moderately high (of the order of $10^3$--\mbox{$10^4$\,cm$^{-3}$}), not typical of interstellar clouds, and thus reinforcing a possible link to the circumstellar environment of the hot star. Kinetic temperatures are around 20~K, consistent with the distance to the star and slightly above typical values of the diffuse galactic interstellar medium. 

The total molecular masses quoted in Table \ref{coldens}(a) were computed after the integration of $N$(CO) over the solid angles of the three regions, assuming a distance of \mbox{10.1\,kpc}. We also considered a factor of 1.36 to account for the helium contribution to the molecular mass. The components A and C have masses of the order of one solar mass, which is also compatible with a stellar origin. Contrarily, component B has 78\msun, almost impossible to be obtained only as a result of mass ejection. The mass derived for component B reinforces the origin suggested in subsection \ref{subsec:relationship}.

The very low [CO/$^{13}$CO] ratio is outstanding. It is in the range $\approx 2$ to 10, far from the solar vicinity value of 89 \citepads[][and references therein]{2004ARA&A..42...39C}. It is well known that intermediate-mass stars like asymptotic giant branch (AGB) stars reach values in the range 20--30 \citepads{2017A&A...599A..39A}, probably due to extra-mixing processes of CNO by-products, added to the first dredge-up \citepads{2010ARNPS..60..381W}. However, the ratio measured in \mbox{AFGL\,2298} is only comparable to those found in evolved massive stars like the red supergiant \object{$\alpha$ Ori} \citepads{1984ApJ...284..223L}, the yellow hypergiant \object{IRC+10420} \citepads{2016A&A...592A..51Q}, or the LBVs \object{MN101} \citepads{2019MNRAS.482.1651B} and \mbox{$\eta$\,Car} \citepads{2012ApJ...749L...4L,2020MNRAS.499.5269G}. 
The reaction times of the CNO cycle depend on exponentially with the temperature, thus it is unlikely that the same mixing mechanisms operate in these very hot stars \citepads{2010A&A...517A..38P}. Therefore, the dramatic changes in the survival times of isotopes and the dominant mixing mechanisms would explain the very different ratios observationally measured. Therefore, it is possible that the unusually low [CO/$^{13}$CO] is the consequence of an overproduction of $^{13}$C during the CNO cycle, and/or a dominant destruction of $^{12}$C to form $^{14}$N \citepads{2010ARNPS..60..381W}.

\subsubsection{Other molecules}

For the remaining molecules, most of which were only observed at 3 mm, we followed a different approach. Being likely less abundant and optically thin, we estimated their respective column densities under LTE conditions, using the software package \texttt{MADCUBA}. In particular, we used the task \texttt{SLIM-AUTOFIT} \citep{2019A&A...631A.159M}, which performs a nonlinear least-squares fit to one or more observed spectra of a given species. \texttt{MADCUBA} is also able to deal with molecules with hyperfine splitting, such as C$^{17}$O, CN, HCN, N$_2$H$^+$, and C$_2$H. Using initial guesses from Gaussian fittings to the lines, convergence was reached in all cases, with the only assumption of an excitation temperature of 20~K.

The column densities obtained by this method are shown in Table \ref{coldens}(b). Besides C$^{17}$O, the largest values are found in C$_2$H and CN, with column densities of the order of \mbox{$10^{13}$\,cm$^{-2}$}. Contrarily, the lowest values are from N$_2$H$^+$, which has only \mbox{$9\times 10^{11}$\,cm$^{-2}$} in the component A. None of these molecules have been detected in component C, not even in the stacked spectra. 

\subsection{Abundances}

To broadly compare \mbox{AFGL\,2298} with other astrophysical environments, and also to investigate possible chemical differences between components, we translated the column densities in Table \ref{coldens} into fractional abundances relative to H$_2$. Since we lack direct measurements of $N(\mathrm{H_2})$ towards \mbox{AFGL\,2298}, we assumed a CO/H$_2$ abundance of $8\times10^{-5}$ \citepads{1975ApJ...202...50D}. The results are listed in Table \ref{abundances}. We remark, however, that the method described in the previous section only provides mean values of the column densities and assumes that all the molecular emission arises in the same volume and comparable depths. This is not expected to strictly occur because the less abundant molecules have usually higher critical densities and emit mostly from deeper regions than CO. Therefore, the abundances of the detected molecules should be regarded as averaged over the whole regions, and eventually considered as lower limits of the real abundances in the volumes they emit.

Components A and B seem chemically different, with fractional abundances in component B being systematically lower than in component A by one or even two orders of magnitude. The abundances of some of the molecules in component B are not far from those measured in cold clouds \citepads[see, for example,][]{1998ApJ...503..717H,2015ApJS..219....2J}.

\begin{figure}
\centering
\includegraphics[width=0.95\columnwidth]{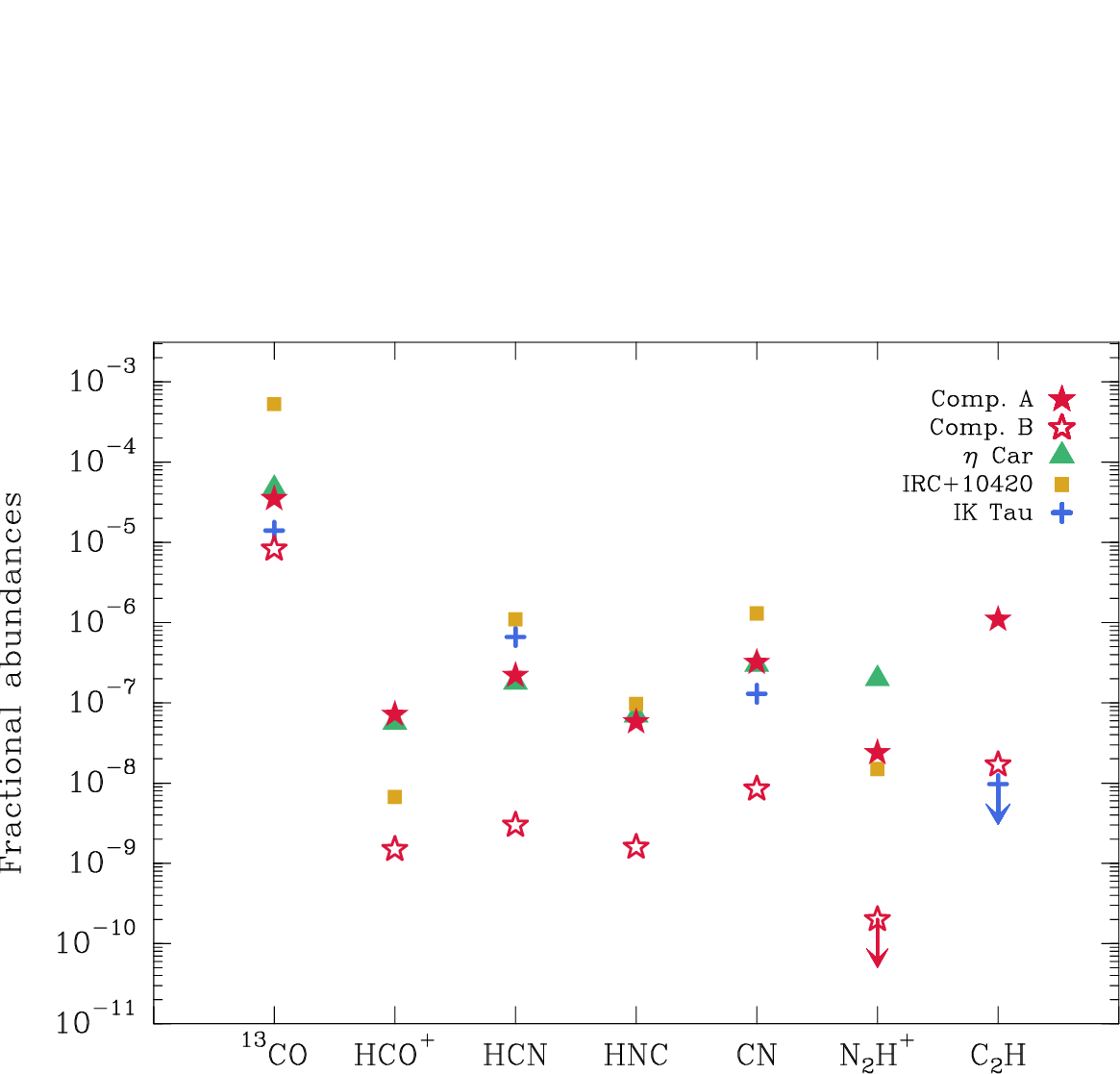}
\caption{Fractional abundances of $^{13}$CO, HCO$^+$, HCN, HNC, CN, N$_2$H$^+$ and C$_2$H in \mbox{AFGL\,2298} compared to other evolved stars. Abundances taken from \citet{2012ApJ...749L...4L} for \mbox{$\eta$\,Car}, \citet{2016A&A...592A..51Q} for IRC+10420, and \citep{2017A&A...597A..25V} for IK\,Tau.
}
\label{fig:abundances}
\end{figure}

Figure \ref{fig:abundances} compares the fractional abundances of the two components with those measured in other well studied evolved stars: the LBV \mbox{$\eta$\,Car} \citepads{2012ApJ...749L...4L}, the yellow hypergiant IRC+10420 \citepads{2016A&A...592A..51Q}, and the oxygen-rich giant \mbox{IK\,Tau} \citep{2017A&A...597A..25V}. While component B has abundances systematically below the values found in all the stars sampled (except the upper limit of C$_2$H in \mbox{IK\,Tau}), the abundances of component A are comparable to those found in the three other stars.

In particular, the abundances of $^{13}$CO, CN, HCO$^+$, HCN and HNC in component A are strikingly similar to those reported in \mbox{$\eta$\,Car} \citepads[see table 2 in][]{2012ApJ...749L...4L}, reaching an agreement better than $\sim$30\% in most cases. The molecular gas in \mbox{$\eta$\,Car} is made of ejecta expelled in the Great Eruption of the 19th century, so such a chemical resemblance may hint that component A is composed of stellar material as well. 

A notable exception is the molecular ion N$_2$H$^+$, which is under abundant in component A with respect to the Homunculus by roughly an order of magnitude, and not detected in component B with an upper limit at least three orders of magnitude below the measured abundance in \mbox{$\eta$\,Car}. The under abundance in \mbox{AFGL\,2298} with respect to \mbox{$\eta$\,Car} may be explained by two destruction processes. For $T_\mathrm{k} < 20$ K, provided that abundant gas-phase CO is available, N$_2$H$^+$ is destroyed producing HCO$^+$ by the reaction \mbox{N$_2$H$^+$ $+$ CO $\rightarrow$ HCO$^+$ + N$_2$} \citepads{2004A&A...416..603J}. The abundance of HCO$^+$ in component A, comparable to that of the Homunculus, might indicate that this destruction path dominates this component. In component B, dissociative recombination with free electrons may drive the destruction of N$_2$H$^+$ through the reaction \mbox{N$_2$H$^+$ $+$ e$^-$ $\rightarrow$ N$_2$ + H}, more efficient at higher temperatures \citepads{2012ApJ...757...34V}. Component B lies in the preferential SE-NW direction where PAHs are detected (see Fig. \ref{vla}, panel b) and where some stratification of the CO isotopologues may be hinted (Fig. \ref{all_comps}, middle row).

This arrangement of molecules and PAHs resembles well studied photodissociation regions (PDRs), like the Orion Bar \citepads{2015A&A...575A..82C} and the Horsehead nebula \citepads{2005A&A...435..885P}, both with fractional abundances of C$_2$H around $10^{-8}$. The overabundant C$_2$H, a prime tracer of PDRs \citepads{2005ApJ...634.1133R}, might be located in an intermediate region between the PAHs and the bulk of component B. Unfortunately, this (admittedly speculative) scenario can not be characterized in this work. The angular resolution of our maps makes it difficult to establish firm conclusions about the chemical stratification and the existence of a PDR.

Taken altogether, these chemical footprints support the hypothesis of component A being the consequence of a relatively recent mass loss event. Conversely, abundances of component B are compatible with interstellar material accumulated by the winds of the central hot star: the HCN and HCO$^+$ abundances, of $3\times10^{-9}$ and $1.5\times10^{-9}$, respectively, are close to the typical values of the ISM and molecular clouds \citepads[see, e.g.,][]{2019A&A...622A..26G, 2019A&A...624A.105F}. This swept-up material would explain the rather large mass of 78\msun\ derived in Sect.~\ref{ndradex}. 

Finally, components A and C may be part of a single mass-loss event, considering the bipolar symmetry displayed (see Fig.~\ref{vla}a). This event would have been in any case a highly non-isotropic mass ejection, in line with the findings of \citetads{2005A&A...437L...1U} and \citetads{2010ApJ...721.1404B}.

\section{Conclusions} \label{sec:conclusions}

We present the first in-depth study of the molecular environment of the Galactic LBV \mbox{AFGL\,2298}, based on observations carried out with the \mbox{IRAM-30\,m} telescope. Our main findings are summarized below:

\begin{itemize}
\item We detected emission from CO and its main isotopologues arising from multiple kinematic components in the observed field. While most of this emission is due to background or foreground molecular clouds unrelated to the source, we identified three velocities of interest where the emission exhibits morpho-kinematic features suggesting a connection with the star, at $\sim$36 (component A), $\sim$46 (B) and $\sim$70 \kms\ (C).
\item Other seven carbon- and nitrogen-bearing molecular species were detected in the field, all of them restricted to the specific velocity ranges of components A and B: HCO$^+$, H$^{13}$CO$^+$, HCN, HNC, CN, N$_2$H$^+$ and C$_2$H. All the lines are present in components A and B, except N$_2$H$^+$ which is not detected in component B.
\item The most intense emission of components A and C appears aligned following a NE-SW line, a preferential direction in which the radio continuum is strongly asymmetric, possibly explained in terms of a shock-ionisation scenario. Likewise, component B, surrounding the dusty nebula from N to SE, seems chemically stratified in the SE-NW direction, where PAH emission appears in the infrared spectrum.
\item Components A and C depict a bipolar symmetry and could be part of a post-main sequence collimated mass-loss event. The morphology and abundances of component B suggest that this component is mainly made of swept-up material from the surrounding cloud.
\item Following a detailed radiative transfer modelling, we determined the average physical conditions of the gas in the three components, along with the column densities of the detected molecules. The gas is moderately dense and close to LTE, with $n(\mathrm{H_2})$ of $10^3$--$10^4$ cm$^{-3}$ and kinetic temperatures of $\sim$20 K, slightly above the typical values for interstellar clouds. Besides, the [$^{12}$CO/$^{13}$CO] ratio in components A and C is significantly lower than in B, which is compatible with heavily processed material. These components could be mostly composed of stellar ejecta, which is also reinforced by their masses (around 1\msun). 
\item The fractional abundances of most of the species match very well with those determined in \mbox{$\eta$\,Car} and the Homunculus nebula from single-dish observations, being compatible with the nitrogen-rich nature of LBV ejecta and further reinforcing the possible stellar origin of some of the gas.

\end{itemize}

These detections are a first step towards a proper characterization of the molecular environment of \mbox{AFGL\,2298}, which now becomes the LBV with the second-largest molecular inventory after \mbox{$\eta$\,Car}. Unfortunately, the limited angular resolution of our data and the non-negligible contamination from extended emission limits the extent of our analysis. Follow-up interferometric observations will properly disclose the relative distribution, kinematics and chemistry of the circumstellar molecular gas, which will help to disclose its true nature and its relationship to the infrared and radio continuum features. 

\begin{acknowledgements}
J.R.R. acknowledges support from grant PID2019-105552RB-C41 funded by MCIN/AEI/10.13039/501100011033. This research has made use of the VizieR catalogue access tool (DOI: 10.26093/cds/vizier).
\end{acknowledgements}

\bibliographystyle{aa}
\bibliography{references.bib}

\begin{appendix}
\onecolumn
\section{Integrated maps of the other molecules} \label{app:other_molec}
In this Appendix, we present the maps of the other molecules, integrated in the velocity ranges of the three components.

%  Component A --------------------------------------------------------
\begin{figure*}[!ht]
%\centering
\sidecaption
\includegraphics[width=10.5cm]{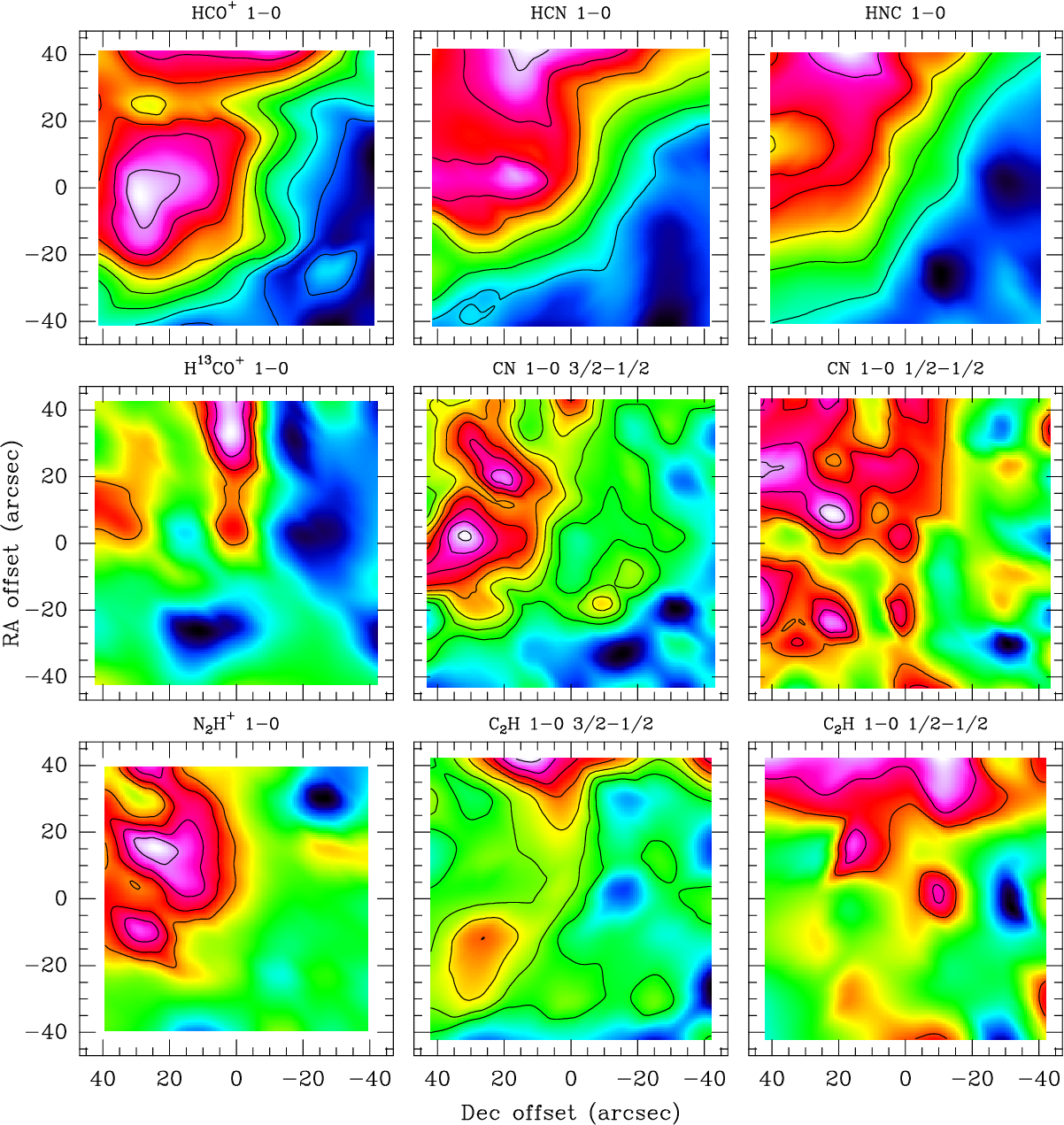}
\caption{Emission of the other molecular lines, integrated in the velocity range of the component A, from 35 to 38\kms. Molecule and transition are indicated on top of each map. First contour and spacing are 0.15 and 0.05 K\,\kms\ in all cases except the top row, where they are 0.3 and 0.15 K\,\kms, respectively.
}
\label{map_A}
\end{figure*}
%

%  Component B --------------------------------------------------------
\begin{figure*}[!hb]
%\centering
\sidecaption
\includegraphics[width=10.5cm]{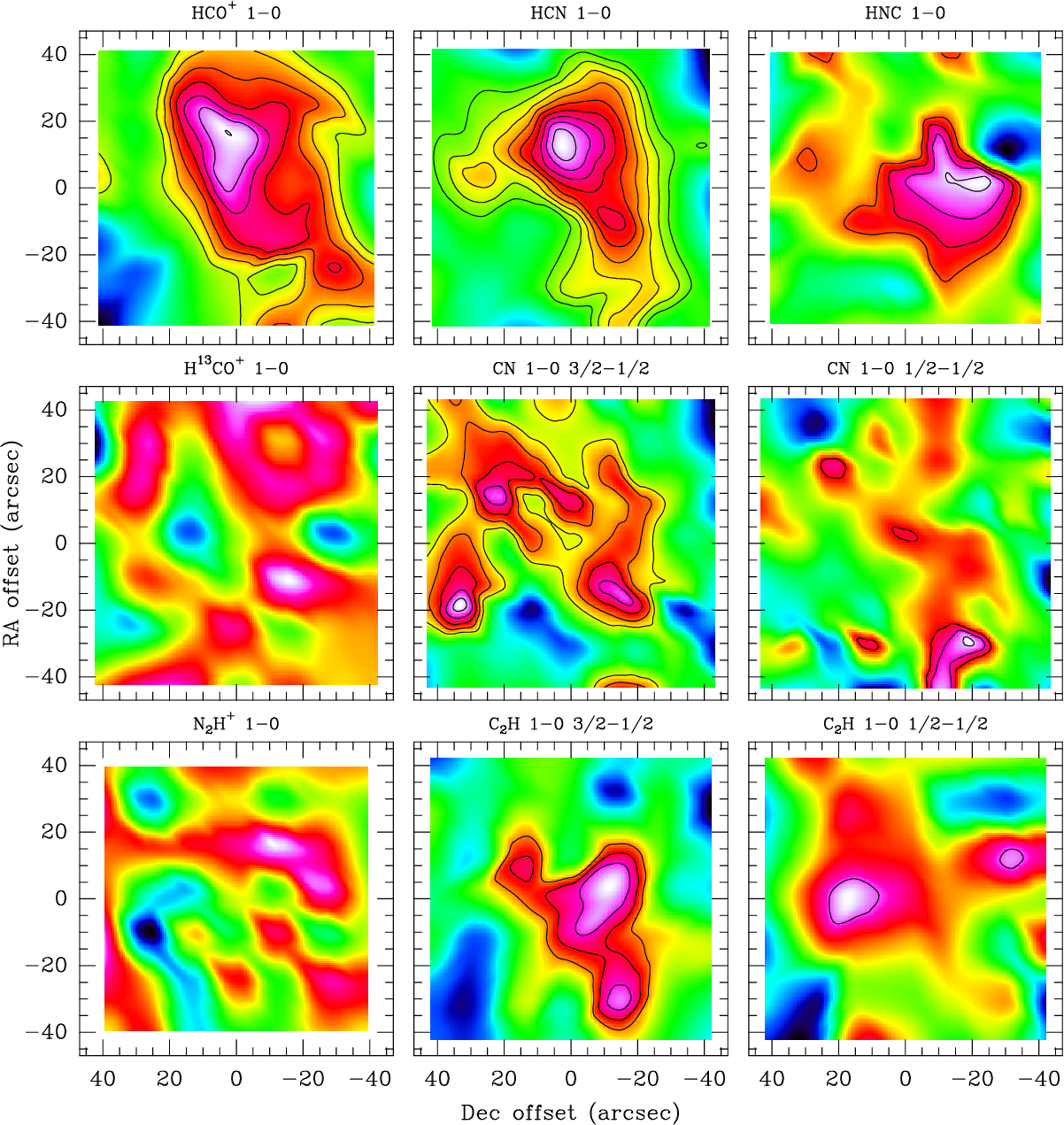}
\caption{Emission of the other molecular lines, integrated in the velocity range of the component B, from 45 to 48\kms. Molecule and transition are indicated on top of each map. First contour and spacing are 0.2 and 0.05 K\,\kms\ in all cases except HCO$^+$, where they are 0.3 and 0.07 K\,\kms, respectively.
}
\label{map_B}
\end{figure*}
%

%  Component C --------------------------------------------------------
\begin{figure*}
%\centering
\sidecaption
\includegraphics[width=10.5cm]{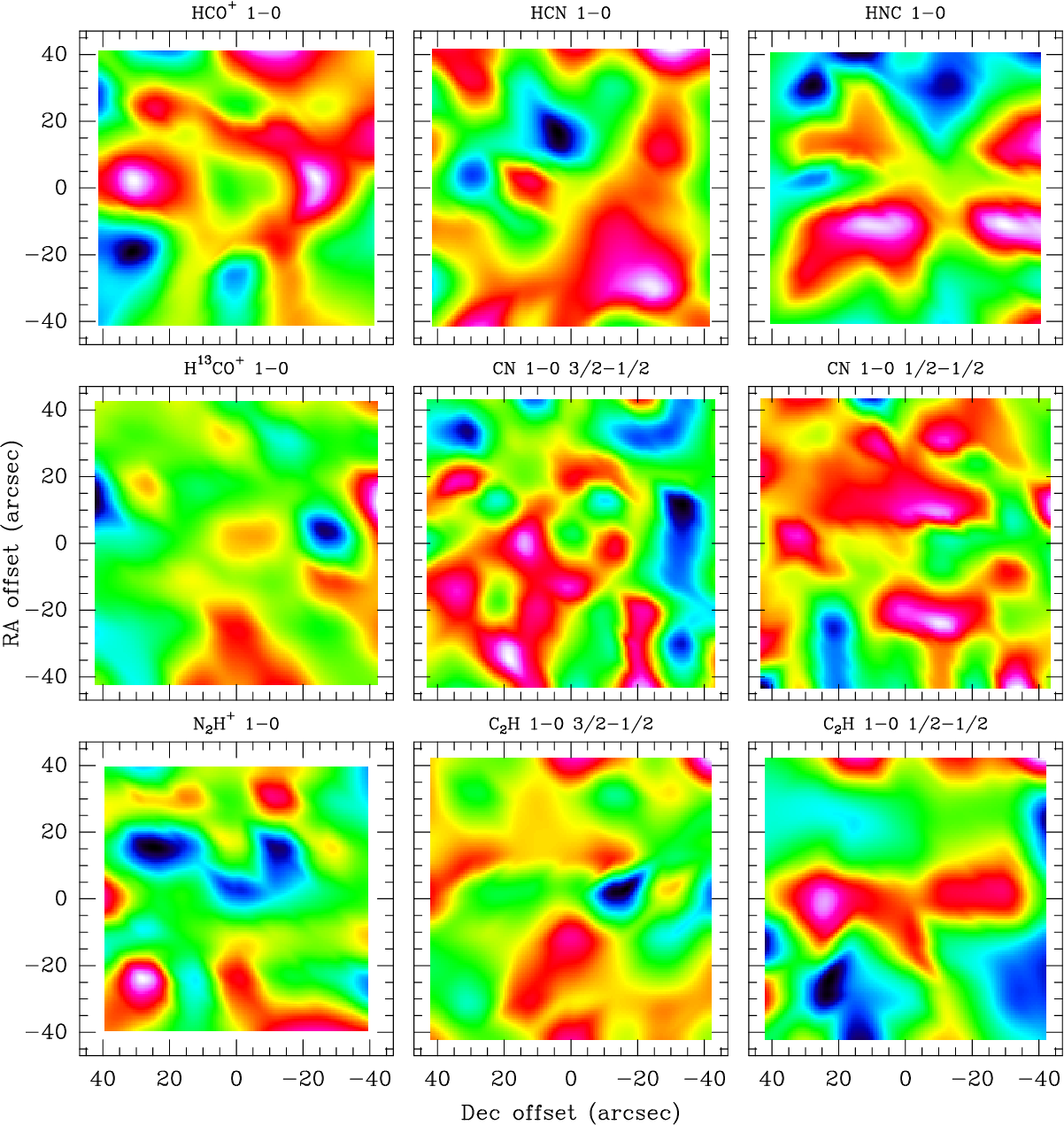}
\caption{Emission of the other molecular lines, integrated in the velocity range of the component C, from 69 to 72\kms. Due to a lack of any significant signal, contour levels are not included.
}
\label{map_C}
\end{figure*}

\end{appendix}

\end{document}